\documentclass[useAMS,usenatbib]{mn2e}
\usepackage{graphicx}

\title[The innermost region of  the NGC~1023 Group]{The innermost region of  the NGC~1023 Group: Insight into its evolution.}
\author[D. Bettoni et al.]{D. Bettoni$^{1}$\thanks{E-mail:
daniela.bettoni@oapd.inaf.it}, L. Buson$^{1}$, P.
Mazzei$^{1}$ and G. Galletta$^{2}$ \\
$^{1}$INAF - Osservatorio Astronomico di Padova\\
$^{2}$Dipartimento di Fisica e Astronomia Universit\'a di Padova}
\begin{document}

\date{Accepted  Received ; in original form;}

\pagerange{\pageref{firstpage}--\pageref{lastpage}} \pubyear{2012}

\maketitle

\label{firstpage}

\begin{abstract}
The NGC~1023 group is one of the most studied nearby groups. We want to give an insight into the evolution of its innermost region by means of ultraviolet observations and proper models. We used the FUV and NUV GALEX archival data as well as a large set of SPH simulations with chemo-photometric implementation. From the UV observations we found that several, already known, dwarf galaxies very close to NGC~1023 are also detected in UV and two more objects (with no optical counterpart) can be added to the group. Using these data we construct exhaustive models to account for their formation. We find that the whole SED of NGC~1023 and its global properties are well matched by a  simulation which provides a minor merger with a companion system 5 times less massive. The strong interaction phase started 7.7\,Gyr ago and the final merger 1.8\,Gyr ago. 
\end{abstract}


\section{Introduction}
The NGC~1023 group is one of the most extensively systems studied (300 related references are listed in the NASA/IPAC Extragalactic Database). The first large scale study of it has been performed by \citet{tully}. Later on the presence of several outstanding 
dwarf galaxies drew the attention of \citet{davies} who studied in detail four of them in the optical bands. 

The most peculiar feature of
NGC~1023 is its proximity to the (likely tidally interacting) fainter galaxy NGC~1023A.  The latter galaxy appears as a small companion located at the East end of NGC 1023. It appears as a low-luminosity condensation partially embedded in the disc of the larger galaxy. This configuration gives the appearance of an asymmetric spiral arm and led \citet{arp66} to include the two galaxies in his Atlas of Peculiar Galaxies as Arp 135. Only later \citet{BC} proposed NGC~1023A to be an interacting individual galaxy stressing also its coincidence with a cloud of a neutral hydrogen (HI) identified by \citet{renzo}. Additional, more recent HI observations confirming the
Sancisi's results (presence of tails and bridges) has been performed by Morganti et al. (2006). The separation (and different stellar population) of the two objects is assured today also by the UV images of the GALEX satellite (Morrissey et al. 2007). In particular NGC~1023A appears to be an irregular galaxy close to the SB0 NGC~1023. Moreover the surface brightness, FUV and NUV GALEX profiles of NGC~1023 are included in the Ultraviolet Atlas of Nearby Galaxies of Gil de Paz et al. (2007).

Current programs (e.g. Trentham \& Tully 2009) are intended to explore the NGC~1023
environment in order to detect the faintest candidate dwarfs (and establish their possible membership to the group);  the outcome of such research could indeed help to clarify, for instance, the so called fundamental missing satellite problem.  Moreover NGC~1023 looks to be the first known S0 for which a proper kinematical approach based on PNe (Cortesi et al. 2011) has shown that it likely originates from the secular evolution of a parent spiral galaxy instead of a more disruptive fast merger.

In this paper we analyze UV data in the central region of the group and we use them to characterize its core with SPH simulations. These are used also to investigate the dark matter distribution in the group environment as not confined in the haloes of the single galaxies.
\begin{figure}
\centering
\includegraphics[width=7.0cm,height=11.0cm]{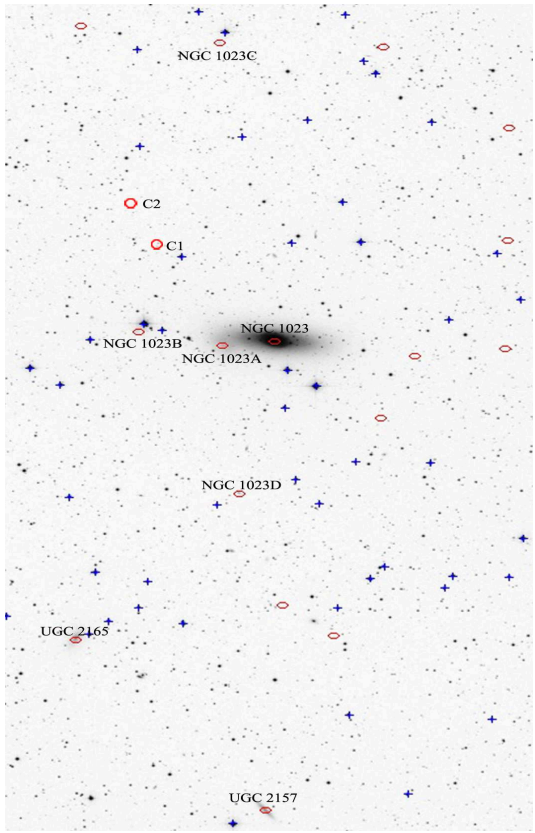}
\caption{GALEX Identification of the galaxies belonging to the centre of the group. It includes known galaxies from Davies \& Kinman (1984) and \citet{tt} and newly discovered UV-bright fainter blobs identified as C1 and C2.}
\label{galex_field}
\end{figure}
The work is structured as follows: section~2 gives a detailed description of the dominant galaxy, NGC~1023, while section~3 provides
an exhaustive description of the UV photometric analysis of both NGC~1023 and NGC~1023A. The adopted SPH modeling is described in section~4. The conclusions are finally listed in section~5.

\section{Nature of NGC~1023}

NGC~1023 is an almost edge-on lenticular galaxy, classified as SB0 by most catalogues (e.g. RC3; De Vaucouleurs
et al. 1991), at an estimated distance of  11.4~Mpc (Tonry et al. 2001). It  belongs to a sparse group, rich in spirals, and is the brightest member of the LGG 70 group (Garcia 1993).  In the optical images (The Carnegie Atlas 1994) the galaxy appears to be composed of a faint stellar disk, a inner and brighter lens and a bar at $\sim$45$^\circ$ from the apparent disk major axis. A central nuclear disk is also present, with relatively young stars (cf. Debattista et al. 2002). From optical line index measurements (Boroson et al. 2011, their Table 1) derive a velocity dispersion of 210\,km/s and a stellar  age of 4.7 Gyr.

The redshift difference with the companion NGC~1023A is $\Delta$V=127$\pm$30 km s$^{-1}$  (Capaccioli et al. 1986) and the projected distance $\sim$7.6~kpc (Barbon \& Capaccioli 1975; adopted H$_0$ = 75 km s$^{-1}$ Mpc$^{-1}$). In  bluer and deeper images the smaller galaxy exhibits some luminous knots that are consistent with a late-type dwarf galaxy.
\begin{figure*}
\centering
\includegraphics[]{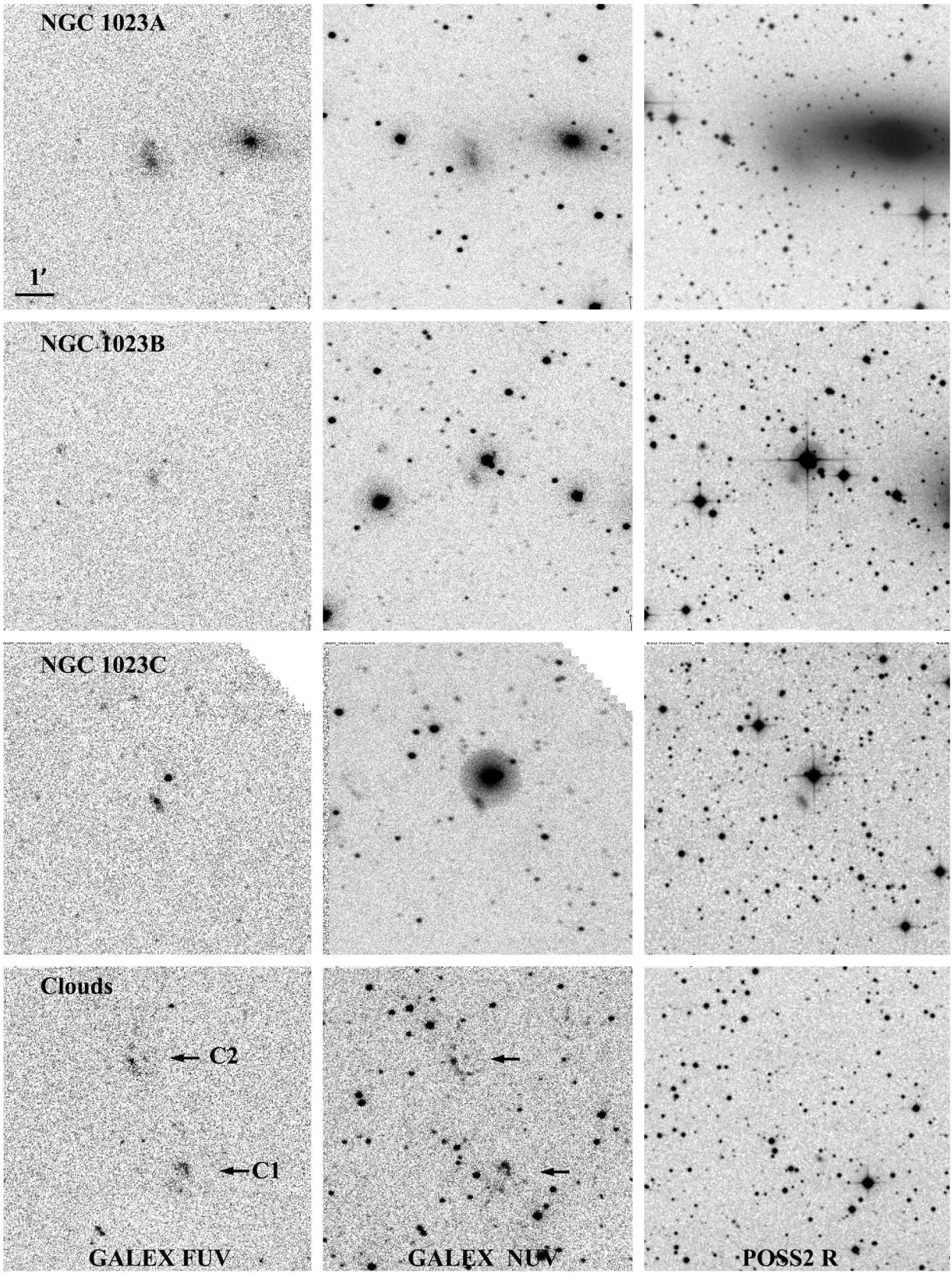}
\caption{Images of the galaxies belonging to the centre of the group. This includes all galaxies from Davies \& Kinman (1984), the galaxies in \citet{tt} here detected in FUV or NUV and the newly discovered UV-bright fainter blobs, C1 and C2.  All panels have the same size with North at the top and East left.}
\label{galex_field1}
\end{figure*}
\begin{figure*}
\centering
\includegraphics[]{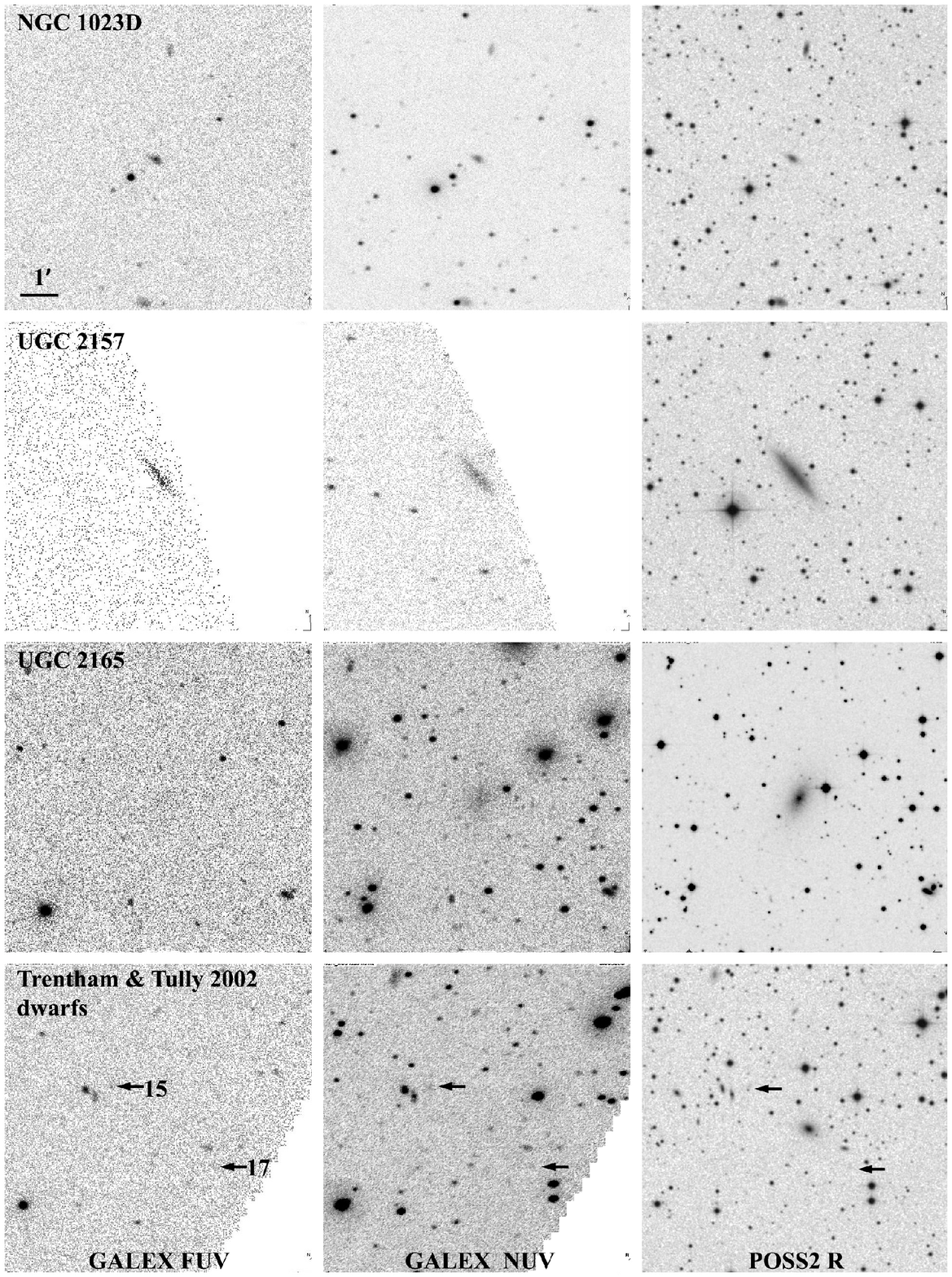}
\caption{Images of the galaxies belonging to the centre of the group. This includes all galaxies from Davies \& Kinman (1984), the galaxies in \citet{tt} here detected in FUV or NUV and the newly discovered UV-bright fainter blobs, C1 and C2. All panels have the same size with North at the top and East left.}
\label{galex_field2}
\end{figure*}
\section{UV photometry}

We used FUV and NUV background-subtracted, de-archived intensity images from the GALEX pipeline.  In order to estimate the uncertainties of UV (AB) magnitudes, we propagated Poisson statistical errors on source and background counts. In addition to the statistical error, we added an uncertainty to account for systematic inaccuracies in the zero-point of the absolute calibration of 0.05 and 0.03 mag for FUV and NUV respectively (Morrissey et al. 2007). 

FUV and NUV magnitudes have been computed as 
$$ m(AB)_{UV} = -2.5 \times log CR_{UV} + ZP $$
where CR is the dead-time corrected, flat fielded count rate, and the zero points are ZP=18.82 and ZP=20.08 in FUV and NUV respectively (Morrissey et al. 2007). We corrected UV magnitudes by adopting a foreground reddening from Schlegel et al. (1998) for both NGC~1023 and NGC~1023A, namely E(B-V)=0.061. With reference to the extinction curve of Savage \& Mathis (1979), this translates into the extinction values A$_{NUV}$=0.56 and  A$_{FUV}$=0.49  respectively, adopting $\lambda_{eff}$=2267~\AA~ and 1516~\AA~for the two above filters. 

For our study, first we derive the photometic parameters of NGC 1023 and NGC 1023A and as a second step we cross-correlate the \citet{tt} sample of objects with the GALEX photometric catalogue. We found 9 matches (together with NGC 1023 and NGC 1023A) over 28 objects. All the galaxies with know redshift are detected in UV bands while only two (namely NGC1023-15 and NGC1023-17) of the dwarfs listed in \citet{tt} have UV emission. In addition to these galaxies we discovered two new objects, identified as C1 and C2. Our GALEX identification of the galaxies, as listed in \citet{davies} and \citet{tt}, are shown in Fig. \ref{galex_field}. Both FUV and NUV surface photometry have been performed using IRAF\footnote{IRAF is written and supported by the IRAF programming group at the National Optical Astronomy Observatories (NOAO). NOAO is operated by the  Association of Universities for Research in Astronomy (AURA), Inc. under cooperative agreement with the National Science Foundation.} STSDAS ELLIPSE routine. ELLIPSE computes a Fourier expansion for each successive isophote (Jedrzejewski 1987), resulting in the surface photometric profiles. 

\begin{figure}
\centering
\includegraphics[width=8.5cm,height=4.0cm]{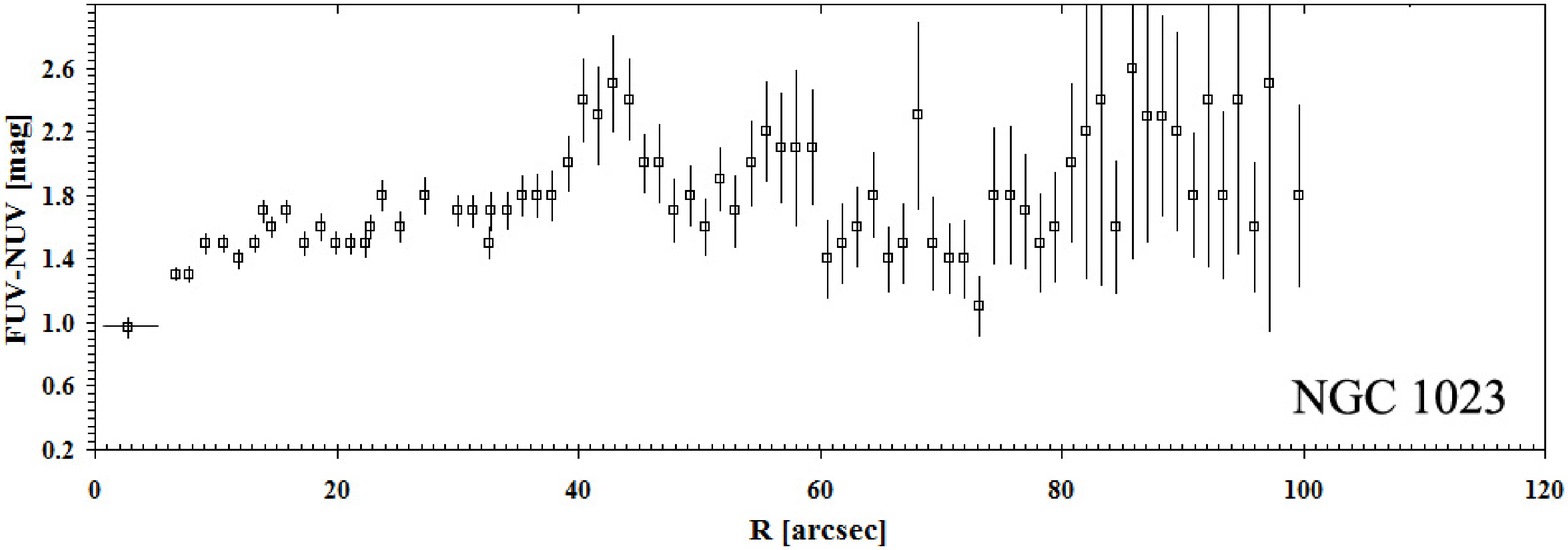}
\includegraphics[width=8.5cm,height=4.0cm]{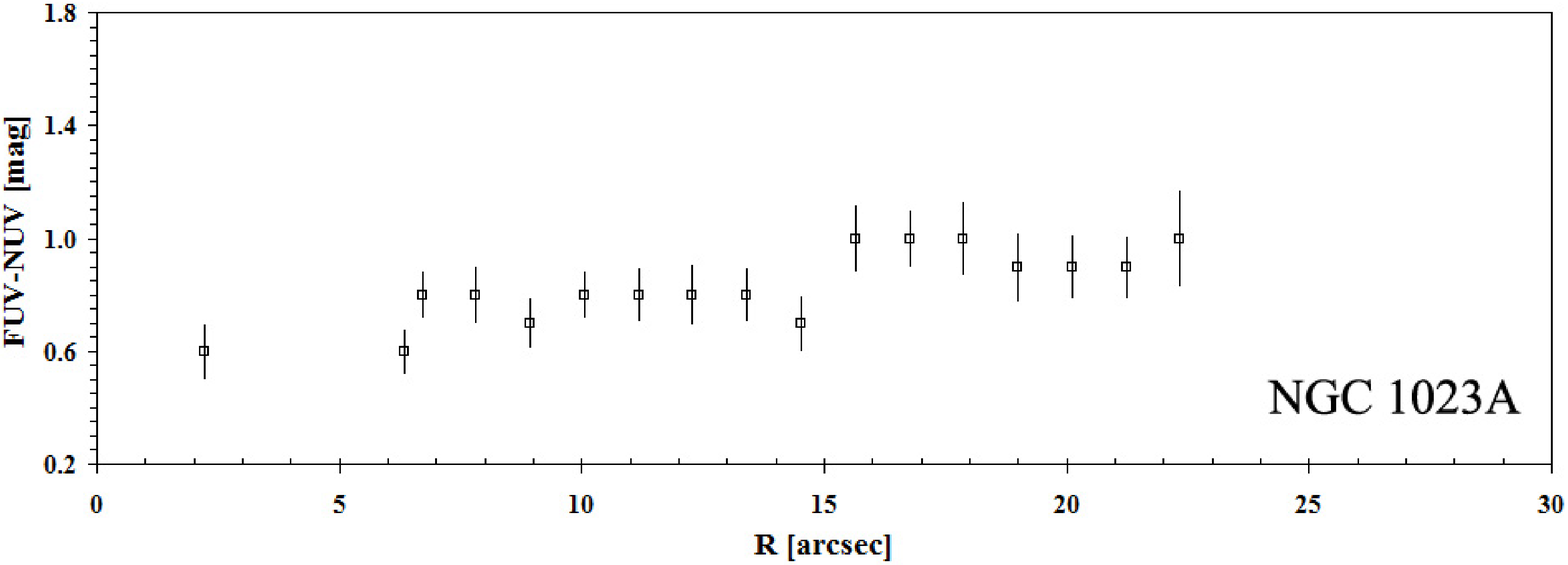}
\caption{FUV-NUV colors for NGC 1023 and NGC~1023A . Note in the lower panel the bluer (FUV-NUV) color of the whole galaxy body, when compared to the dominant galaxy NGC~1023.}
\label{col1}
\end{figure}

All corrected magnitudes m$_{AB}$s' of NGC~1023, NGC~1023A, and Davies \& Kinman's and \citet{tt} objects are given in Table \ref{tab:fos_indb}. In Figures \ref{galex_field1},\ref{galex_field2} we show the NUV, FUV and POSS2 R images of all the objects we have detected (in NUV or FUV). We computed the ultraviolet integrated photometry (FUV$+$NUV) of each UV-bright source and the color profiles of the two brightest galaxies, namely NGC 1023 and NGC 1023A (Fig \ref{col1}).  
Integrated photometric properties are given in Table \ref{tab:fos_indb} while photometric parameters for the fits are listed in Table \ref{re}. Errors in total magnitudes were obtained from the correlation coefficient of the model fit vs. the observed values.

\begin{figure}
\centering
\includegraphics[width=8.5cm,height=6.0cm]{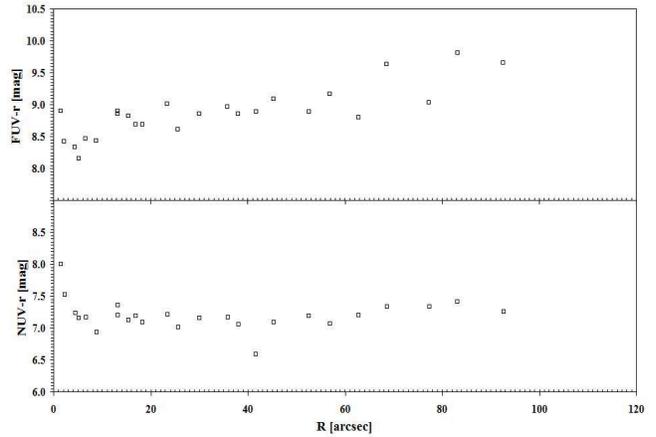}
\caption{NUV-R and FUV-R color profiles for NGC 1023. Note in the upper panel the much red (FUV-R) color of external regions of the dominant galaxy NGC~1023.}
\label{col}
\end{figure}

We fitted the luminosity profiles of NGC~1023 assuming two-components: a bulge following an r$^{1/4}$ law, and an exponential disc. The inner regions of the profiles, affected by the GALEX PSF ($\sim$ 5$"$), were excluded in performing interpolation. The lens and bar are not visible in UV images and the ellipse interpolations find constant ellipticity and P.A. values. In the case of NGC 1023 they are $\epsilon\sim$0.23 (NUV) and 0.12 (FUV); P.A$\sim$80$^\circ$ (NUV) and   83$^\circ$ (FUV). The values of the bulge effective radii  in FUV and NUV reported in Table \ref{tab:fos_indb} are quite similar and are in agreement with those found by optical photometry in R band, reported in the work of \citet{Noor05}: r$_e(bulge)$=32". The stellar disk, when observed in red light, appears larger, with an r$_e(disk)$=62" compared to values lower than 50" found in the UV.   The FUV-NUV color profile (top panel of Fig. \ref{col1}) indicate that the nucleus is bluer that the external regions. This is confirmed also by the NUV-r color profile plotted in Fig. \ref{col} together with the FUV-r. The r data were derived from \citet{Noor05}. 

Both UV profile components can be fitted in NGC~1023A with an exponential disk. The routine ELLIPSE gives almost constant  ellipticity and position angles. They are: $\epsilon \sim$0.45 (NUV) and 0.38 (FUV), and P.A$\sim$16$^\circ$ (both in NUV and  FUV). The effective radii and surface brightness are listed in Table \ref{re} together with the total magnitudes of both galaxies.  Integrated photometric properties are given in  Table \ref{tab:fos_indb} while photometric parameters for the fits are listed in Table \ref{re}.  Contrary to what observed in NGC 1023 the color profile in NGC 1023A is constant around a value $\sim$0.8.

\begin{figure}
\includegraphics[width=8.7cm,height=8.7cm]{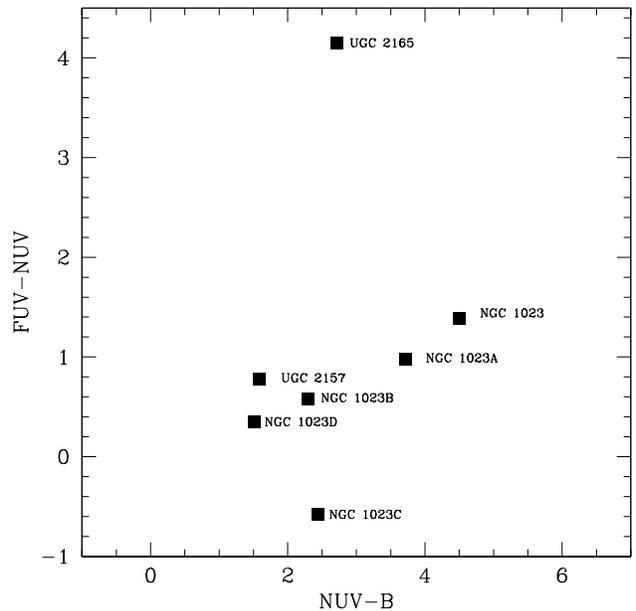}
\caption{UV-optical two-color diagram of the main dwarf galaxies of the NGC~1023 group.}
\label{colori}
\end{figure}
 We measured the UV magnitudes of the two UV emission regions, labeled C1 and C2 (see Fig. \ref{galex_field}, \ref{galex_field1}, \ref{galex_field2}) inside an aperture of r=15 arcsec; the values are listed in Table \ref{tab:fos_indb}. In Fig. \ref{colori} we plot the colors of the main objects in the group. We note that NGC~1023 and NGC~1023A have similar colors, whereas the "dwarfs" NGC~1023B, NGC~1023D, UGC 2157 and especially NGC~1023C are quite bluer in the UV (both in FUV-NUV and NUV-B). UGC 2165 on the contrary is very red, in particular the FUV flux measured can be considered as an upper limit and may indicate a very old stellar population.
\begin{table*}
\caption{Integrated photometric properties}
\begin{tabular}{lcccccccccc}
\hline
Object & R.A. & DEC & {m$_{NUV}$(AB)} & {m$_{FUV}$(AB)} & B & R & FUV-NUV & NUV-B & {SFR$_{FUV}$} &Radial Vel.\\
& (2000) & (2000) & & & mag & mag & mag & mag & $10^{-2}$ M$_{\odot}$ yr$^{-1}$& km s$^{-1}$\\
\hline
\hline
NGC~1023   & 40.099   & 39.065 & 14.85$\pm$0.09  & 16.24$\pm$0.23  & 10.35 & 9.14 & 1.39 & 4.50 & 2.68 & 637\\
NGC~1023A & 40.155   & 39.056 & 16.66$\pm$0.40  & 17.64$\pm$0.38  & 12.95 & 15.30 & 0.98 & 3.71 & 0.73 & 743\\
NGC~1023B & 40.251   & 39.073 & 19.34$\pm$0.20  & 19.92$\pm$0.3   & 17.04 & 17.34 & 0.58 & 2.3 & 0.09 & 593 \\
NGC~1023C & 40.165   & 39.380 & 19.54$\pm$0.16 & 18.96$\pm$0.32 & 17.10 & 15.13 & -0.58 & 2.44 & 0.22 & 903 \\
NGC~1023D & 40.137   & 38.900 & 18.43$\pm$0.22  & 18.78$\pm$0.28   & 16.92 & 16.37 & 0.35 & 1.51 & 0.25 & 695 \\
C1                & 40.230   & 39.161 & 19.49$\pm$0.24  & 19.36$\pm$0.32 &  & & & & 0.15 & \\
C2                & 40.258   & 39.206 & 19.58$\pm$0.3  & 19.54$\pm$0.3 &  & & & & 0.13 & \\
UGC~2165 & 40.314 & 38.744  & 18.01$\pm$0.16 & 22.16$\pm$0.2 & 15.29 & 13.36   &  4.15 & 2.72 & 0.01& 740\\
UGC~2157 & 40.105 & 38.563 & 16.61$\pm$0.038 & 17.39$\pm$0.072 & 15.02 & 13.67 &  0.78  &   1.59 & 0.92 & 488\\
N1023-15 & 40.087 & 38.783 & 19.54$\pm$0.31 & 20.10$\pm$0.25 &  & 19.71 & 0.56 &   & 0.07 &\\
N1023-17 & 40.031 & 38.748 & 20.26$\pm$0.33 & 20.71$\pm$0.12 &  & 20.23 &  0.45 & & 0.04  &\\   
\hline
\end{tabular}
\label{tab:fos_indb}
\small{Note: B magnitudes of NGC~1023A, NGC~1023B, and NGC~1023D are from Davies \& Kinman (1984) and for NGC~1023 from NED, the R magnitudes are from Trentham \& Tully (2002) }\\ 
\end{table*}

\begin{table*}
\caption{Photometric parameters from the fit of the luminosity profile}
\label{Photometric parameters}
\begin{tabular}{lcccccc}
\hline\hline
Object & Band & $m_{tot}$  & $r_{e~(bulge)}$ & $\mu_{e~(bulge)}$ & $r_{e~(disk)}$ & $\mu_{e~(disk)}$\\
& & mag   & arcsec & $mag/"^2$ & arcsec & $mag/"^2$\\
\hline
           &           &                          &                      &                           &                          &                          \\
N1023 & NUV    & 14.85$\pm$0.09 & 34.0$\pm$9.5 & 26.34$\pm$0.38  & 50.00$\pm$1.45 & 27.24$\pm$0.07 \\
N1023 & FUV    & 16.24$\pm$0.23 & 32.0$\pm$6.6 & 27.44$\pm$0.29  & 46.00$\pm$1.45  & 28.84$\pm$0.09 \\
N1023A & NUV & 16.66$\pm$0.40 & --                   & --                       & 25.00$\pm$2.05  & 26.34$\pm$0.08\\
N1023A & FUV  & 17.64$\pm$0.38 & --                   & --                       & 20.0$\pm$1.55   & 26.84$\pm$0.11 \\
           &           &                          &                      &                           &                          &                          \\
\hline
\label{re}
\end{tabular}
\end{table*}

\subsection{Star formation rate}
The present-day Star Formation Rate (SFR) of each UV source can be derived---following 
\citet{Kenn}---using its UV continuum luminosity and the relation\\

 SFR$_{FUV}$ (M$_\odot$ yr$^{-1}$) = 1.4 $\times$ 10$^{-28}$L$_{FUV}$(ergs s$^{-1}$ Hz$^{-1}$).\\
 
 \noindent
Such a relation holds for a  Sapleter's IMF with lower and upper  mass limits 0.1\,M$_\odot$ and  100\,M$_\odot$ respectively. We used our total FUV magnitudes to derive the SFRs which are
given in Table \ref{tab:fos_indb}. Using the previous equation we found that the SFR is not null in NGC 1023  although it provides very low values (2.68$\times 10^{-2}$\,M$_\odot$ yr$^{-1}$). The Spitzer data for the inner regions of this galaxy \citep{shap} indicate no signature of SF. The SFR in the other dwarf galaxies and in the two blobs (C1 and C2) is even lower, with the lowest value, 0.01$\times 10^{-2}$\,M$_\odot$ yr$^{-1}$, for UGC 2165. 
However their colors (see Fig. \ref{colori}) are bluer than those of the two brighter galaxies of the group.

\section{Modeling}

With the aim to match the overall SEDs and the global properties of NGC1023 and NGC1203A in a consistent way with the dynamical properties of the group, we select, from a large set of SPH simulations, one case which gives us insights into the evolution of such a system and  the group itself.

Our SPH simulations of galaxy formation and evolution are starting from the same initial conditions as described in \citet{paola5} (MC03 in the following) and \citet{paola1} i.e., collapsing  triaxial systems initially composed of dark matter (DM) and gas in  different proportions and different total masses.

All the simulations  include self--gravity of gas, stars and DM, 
radiative cooling, hydrodynamical pressure, shock heating, 
artificial viscosity, star formation (SF) and  feedback from 
evolved stars and type II SNe, and  chemical enrichment as in MC03 (and references therein).

They provide the synthetic SED 
at each evolutionary step. The SED accounted for chemical 
evolution, stellar  emission, internal extinction and 
re-emission by dust in a self-consistent way, 
as described in \cite{marilena} and \cite{dani}; 
this extends over four order of magnitude in wavelength,
 i.e., from 0.1 to 1000 $\mu$m. So, each simulation self-consistently 
 provides dynamic, morphological, and chemo-photometric evolution.
The Initial Mass Function (IMF) is of Salpeter type with 
upper mass limit 100$\,M_\odot$ and lower mass limit 0.01$\,M_\odot$ (Salpeter, 1955) (see \citet{CM99} and MC03 for a discussion).
All the model parameters here adopted had been tuned in previous papers  where
the integrated properties of simulated galaxies, i.e., colors, absolute magnitudes, mass to luminosity ratios provided by different choices of model parameters after 15 Gyr, had been successfully compared with those of local galaxies (see also \citet{paola1}; \citet{paolaa}).
In particular, the IMF choice here  provides a slightly higher
SFR compared with the other possibilities examined,  allows for the lowest feedback strength, and the expected rotational support when disk galaxies are formed. Moreover, as pointed out by \citet{K12}, its  slope is almost the same as the Universal Mass Function which links IMF of galaxies and stars to those of brown dwarfs, planets and small bodies (meteoroids, asteroids) \citep{BH07}.

With respect to MC03, the initial particle resolution  here is enhanced to more than 6$\times$10$^4$ (see below) instead of 2$\times$10$^4$, and the
time separation between the snapshots halved,  151\,Myr.

A new large set of galaxy encounters 
involving systems with 1:1 mass ratios and different mass ratios have been performed with the same initial conditions as described in MC03 (and references therein). 
By seeking to exploit a vast range of orbital parameters, we 
carried out different simulations for each couple of interacting 
systems varying the orbital initial conditions in order to have, 
for the ideal Keplerian orbit of two mass points of mass 
equal to 10$^{12}$,  or 10$^{13}\,M_\odot$ , the first peri-center 
separation,  p, equal to the initial length of the major axis of the 
DM triaxial halo, i.e. 88 kpc for 10$^{12}\, M_\odot$, or 
equal to 1/10, 1/5,  and 1/3 of the same axis for 10$^{13}\,M_\odot$ 
encounters. 
For each of these separations, we changed the eccentricity in 
order to have hyperbolic orbits of different energy. The spins 
of the systems are equal (MC03), generally parallel each other, and perpendicular 
to the orbital plane, so we studied direct encounters. However, 
some cases with misaligned  spins have been also analyzed in 
order to deepen the effects of the system initial rotation on 
the results. Moreover, for a given set of encounters with the 
same orbital parameters  we also examined the role of 
increasing initial gas fractions.  All these new simulations will be 
fully discussed elsewhere (Mazzei, in prep).

A simulation able to match the global properties 
 of NGC~1023 is chosen and  discussed below.
All the observables, i.e. absolute magnitudes, colors, which correspond to the SED over almost four order of magnitude in wavelength, and mass-to-light luminosity ratios are self-consistently derived from our simulation at the snapshot selected, and successfully compared with the available data.

\subsubsection{NGC~1023}

The whole SED of NGC~1023 and its global properties are well matched by a  simulation which provides a minor merger.
The total  mass of the primary system is  1$\times$10$^{13}$ M$_\odot$ with gas 
fraction 0.01, that-one of the companion is 5 times less with the same gas fraction.
So, the starting point is given by two triaxial collapsing systems with total mass 1.2$\times$10$^{13}$ M$_\odot$, and  60 times less of gas, i.e. 2$\times$10$^{11}$ M$_\odot$.  The mass particle resolution is 5$\times$10$^{6}$ M$_\odot$ for gas and 4.95$\times$10$^{8}$ M$_\odot$ for DM particles.
This requires 63838 initial particles. The triaxiality ratio, $\tau$ \citep{w92}, is 0.84.\\
The first pericenter separation, 433~Kpc, corresponds to 1/3 of the major axis of the primary system; the orbit eccentricity is 1.3.  and the anomaly corresponds to 200 degrees. 
The spins of the systems are equal ($\lambda$=0.06, MC03), parallel, so we are dealing with a direct encounter, and both aligned with the shorter of their principal axes.\\
Stars are born in the inner regions of their halos after 3\,Gyr from the beginning. Galaxies grow changing their shapes step by step, as their trajectories are approaching and their halos mixing.

\begin{figure}
  \centering
{\includegraphics[width=8cm]{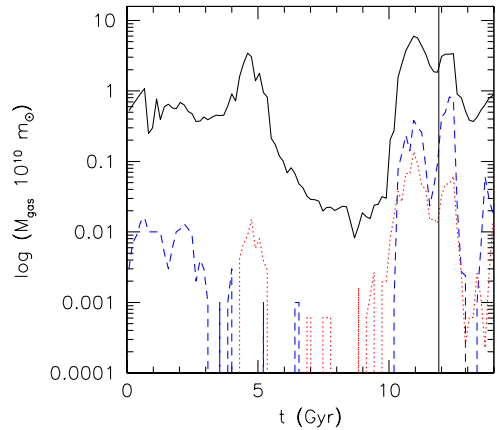}}
      \caption{(Black) continuous line  shows the evolution of the gas mass inside a radius of 50\,Kpc centered on the V band luminous center of the galaxy system; (blue) dashed line corresponds to the gas with temperature $\le$10000\,K, dotted (red) line to the gas with temperature $\ge$  10$^6$\,K. The vertical line emphasizes the age of our fit.}
       \label{accgas}
   \end{figure}
\begin{figure}
  \centering
{\includegraphics[width=9cm]{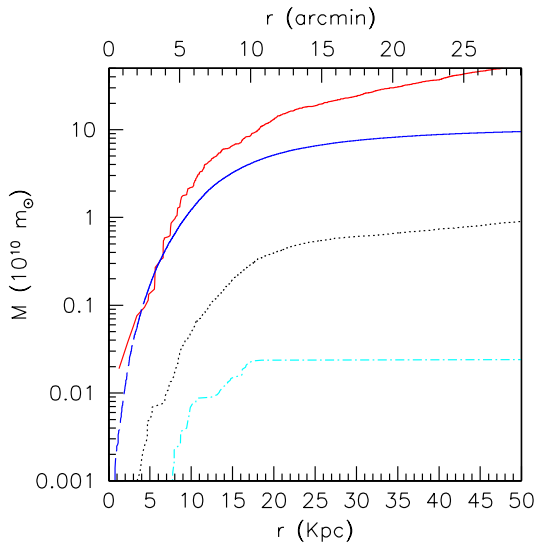}}
      \caption{Mass distribution inside a radius of 50\,Kpc centered on the V band luminous center of the galaxy system; (red) continuous line corresponds to the DM mass, (blue) long dashed line shows that of stars, (black) dotted line gas, and (cyan) dotted-dashed line cold gas, i.e. gas with T$\le$10$^4$\,K. }
       \label{mdist}
   \end{figure}

The fit corresponds to a galaxy age of 11.9 Gyr.
The strong interaction phase between the systems started 7.7\,Gyr ago and the final merger 1.8\,Gyr ago.
Figure \ref{accgas} shows the gas accretion history inside a sphere  centered on the luminous (V band) centre of the  galaxy system with a radius of 50\,Kpc. 
The mass of gas  with temperature $\le$10000\,K, which represents the upper limit of the  cold gas mass inside the selected radius (its cooling time scale is very shorter than the  snapshot time range, 151Myr), is 9.50$\times$10$^8$ m$\odot$ at the age of our fit (the vertical line in Fig. \ref{accgas}).
Such a value well agrees with that derived from eq. (1) of \citet{K85} using D=5.93\,Mpc (NED from 3K CMB) and  FI, the flux integrated over line, 80\,Jy\,km/s  \citep{Noor05}, i.e., 7.3$\times$10$^8$ m$\odot$; this value is at least a factor 3.7 more using the
distance in \citet[their Table 2, 11.4 Mpc]{morg}.
Figure \ref{mdist} shows the mass distribution at the selected snapshot  inside  50\,Kpc, corresponding to 15r$_{eff}$ of NGC~1023.


\begin{figure}
  \centering
 {\includegraphics[width=8cm]{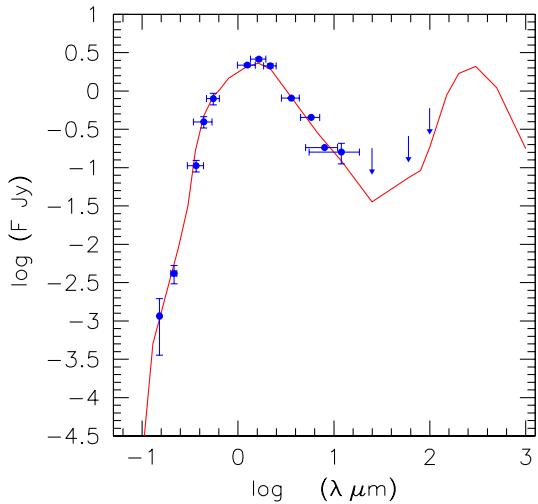}}
      \caption{Continuous line (red) show the prediction of our model (see
      text) for NGC~1023. (Blue) filled circles correspond to  data  in Table 1 for UV spectral range, from NED in the optical-IR range, and from \citet{Paetal04} in IR and FIR. Filled triangles show upper limits; error bars account for band width and 3 \,$\sigma$ uncertainties.
}
       \label{modsed}
   \end{figure}
 The  B band absolute magnitude of the model, $M_B=-20.35$, agrees with the value in Table 1 of \citet{bofa} and with Table 1 of \citet[and references therein, $M_B=-20.42$]{em07};
R band absolute magnitude, $M_R=-21.95$,  agrees with the estimate of \citet{tt}, $M_R=-22.33$ (their Table 2).
Furthermore, all the available data, extending over three order of magnitude in wavelength,  are well matched by the SED provided by our simulation, as shown in Fig. \ref{modsed}.
The FIR region is not strongly constrained by the data, since upper limits of 25, 60 and 100\,$\mu$m fluxes are only available;
dust components (warm and cold with PAH  as discussed in \citet{paola2}) with the same average properties as derived by \citet{paola4} and \citet{paola5}  for a complete sample of nearby early-type galaxies are included (r$_d$=100r$_c$, \citet{paola4}).
\begin{figure}
  \centering                  
 {\includegraphics[width=9cm]{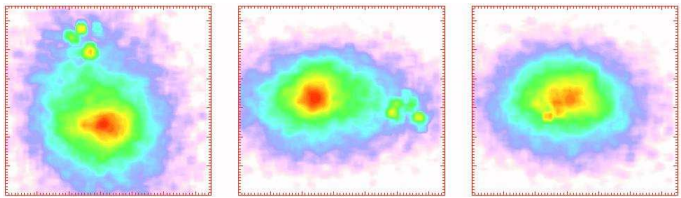}}
      \caption{
 R band contours of XY, YZ and XZ projections of the snapshot that best fits the UV Galex morphology of NGC~1023 (upper panel of Fig~8).
 Every panel corresponds to 50x50Kpc and includes sixty equispaced levels for the same density contrast, equal to 100. 
	}
    \label{morfo_II}
   \end{figure}
Figure \ref{morfo_II} shows R band maps of different projections  of the selected snapshot.

\begin{figure}
  \centering                  
 {\includegraphics[width=9cm]{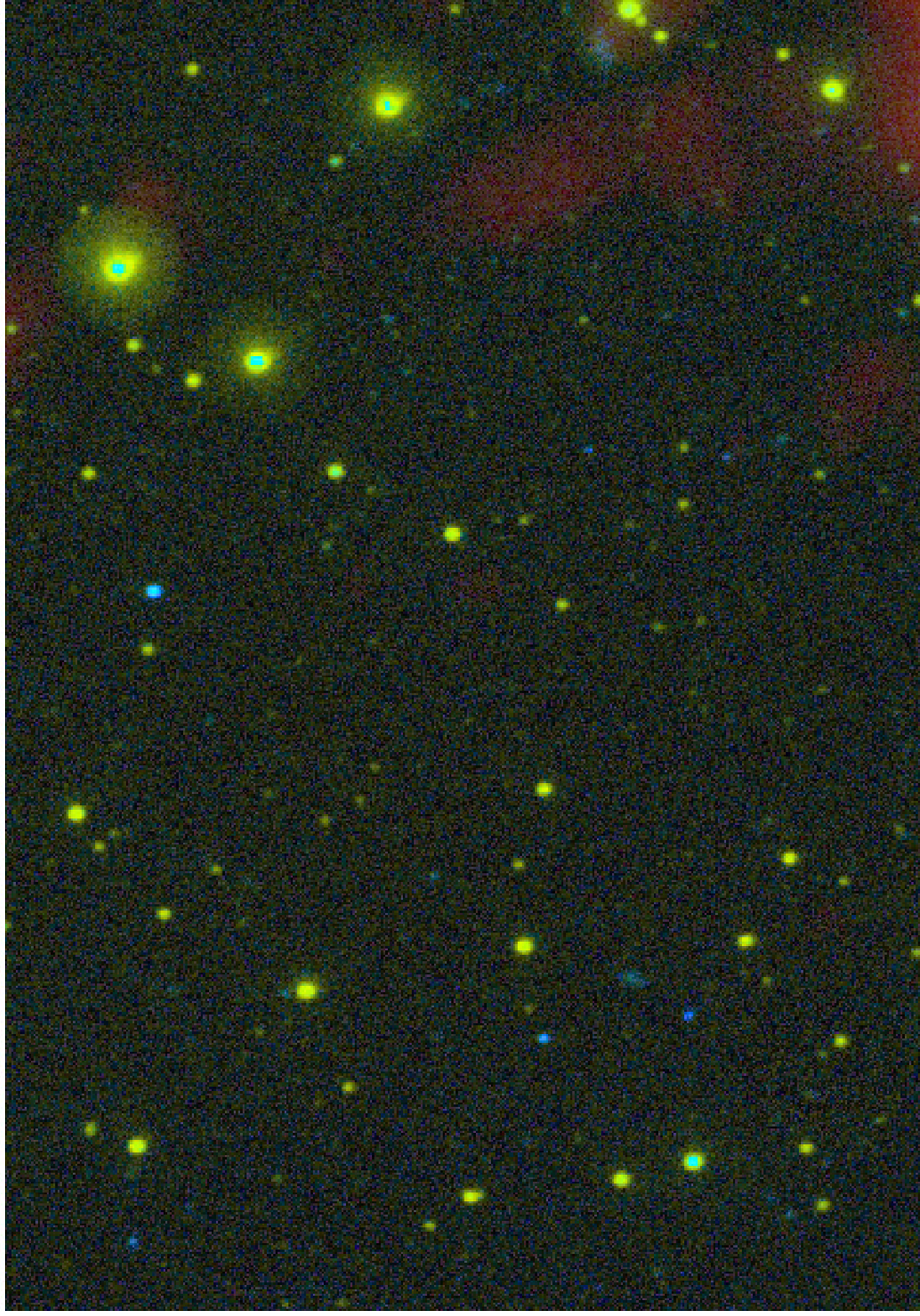}}
 {\includegraphics[width=6.75cm]{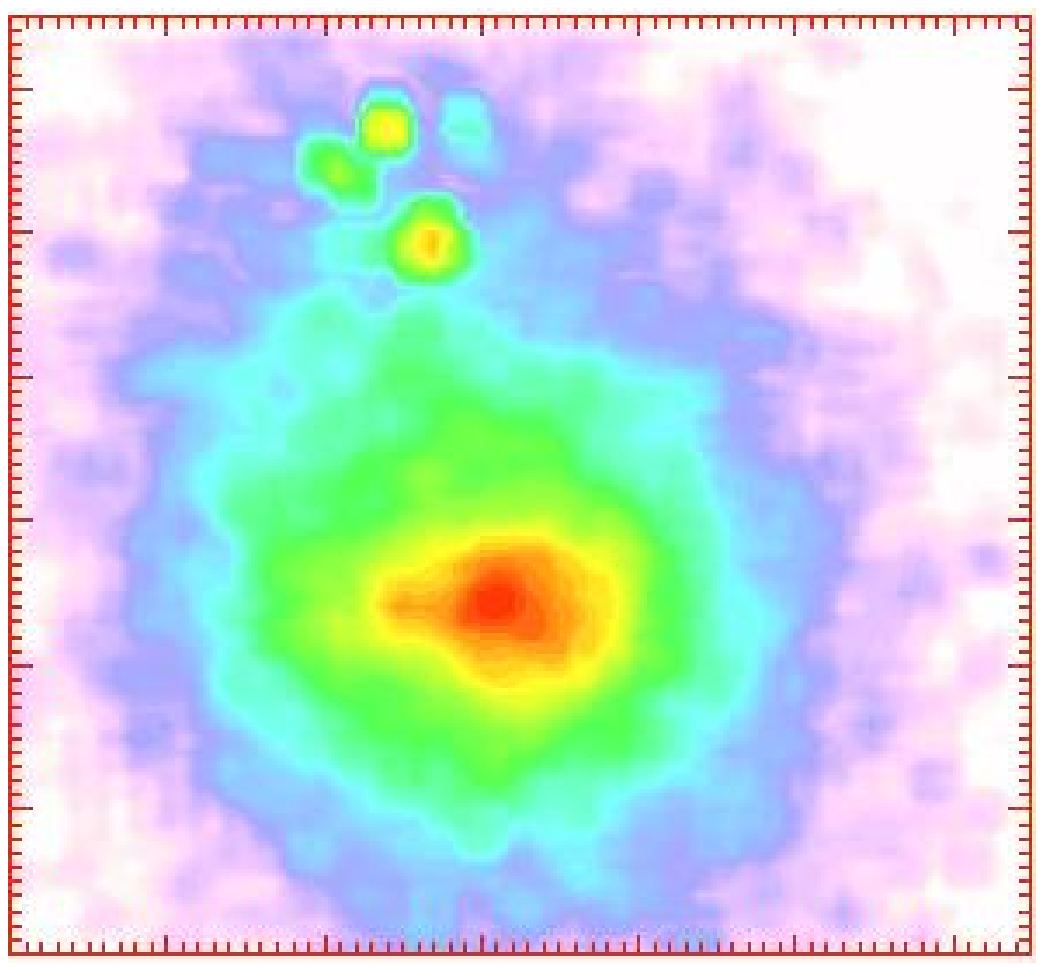}}
      \caption{ HI band image (red) of NGC~1023 superimposed to the UV GALEX image {\sl (top)}, on the same scale (50x50Kpc), with isodensity contours {\sl (bottom)} from our simulation (see text). The simulated image correspond to R band contours of XY projection as in Fig.~7. Note the complex structure of neutral gas in the surroundings of NGC~1023.
	}
    \label{morfo}
   \end{figure}

Figure \ref{morfo} compares, on the same scale, the UV Galex image with superimposed the HI map of NGC~1023, with R map of XY projection as in Fig. \ref{morfo_II}.

Star populations inside the previous maps are  5.0 Gyr old in average, in agreement with \citet{bofa} (see Section ~2). The luminosity weighted age of the model does not depend strongly on the radius nor on the band, it is 2.5-3 Gyr going from the inner regions up to  50\,Kpc. 

From the study of \citet{tt}, the virial mass of the NGC~1023 group  inside 315\,Kpc (their virial/harmonic radius), is  6.4$\pm$3$\times$10$^{12}$\, M$_{\odot}$,  and L$_{R_{\odot}}$=2.0$\times$10$^{10}$\, L$_{\odot}$. We predict a total mass of 5.54$\times$10$^{12}$\, M$_{\odot}$ with a stellar mass of 1.20$\times$10$^{11}$\, M$_{\odot}$,  and L$_{R}$=2.56$\times$10$^{10}$\, L$_{\odot}$ inside the same radius.

The instantaneous SFR in the inner region of the simulated galaxy agrees with
SAURON results by Shapiro et al. (2010) who measure SFR$\sim$0 $M_{\odot}$~yr$^{-1}$ (see Section 3.1), moreover
the total SFR at the selected snapshot, 0.045~M$_{\odot}$~yr$^{-1}$, is in well agreement with
our estimates from FUV luminosity. Note that to compare these values with those in Table \ref{tab:fos_indb} a multiplying factor of $\sim$2.4 has to be applied to the data in such a table to account for the lower mass limit of our simulation.

\begin{figure}
  \centering                  
 {\includegraphics[width=9cm]{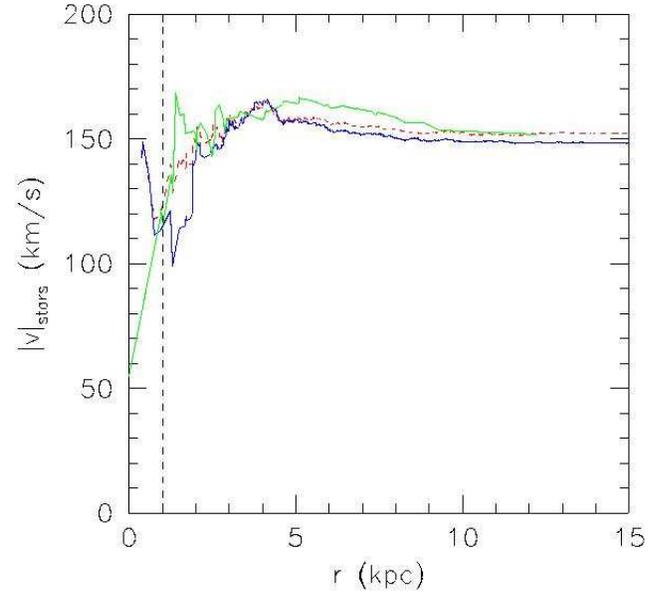}}
      \caption{Star absolute velocity distribution at the selected snapshot. Velocities of star particles are computed in their center of mass (red dashed line) and averaged at each galaxy radius; (blue) continuous line shows the same but weighted on V luminosity.    The green line corresponds to the average velocity distribution of planetary nebulae. Vertical dashed line corresponds to the inner region (R$\leq$1 Kpc) where model sampling could affect the results.}
    \label{vel}
   \end{figure}

Figure \ref{vel}  compares our predicted average velocity distribution of stars at each galaxy radius  with the same  from the data in Fig. 4 of Cortesi et al. (2011).  Velocities from our simulations are higher than 
7\% ; such a reduction factor allow us to account for the inclination angle of NGC~1023 that is $\sim70^{\circ}$. Note that the green line  in Fig. \ref{vel} corresponds to the average velocity distribution of planetary nebulae; we select their velocities along the major axis of such a galaxy (PA=$90^{\circ}$) accounting for a recession velocity of 637\,km/s. 
\section{Conclusions}

First, we reanalyzed the inner group NGC~1023 by means of deep GALEX images. So we extend to UV range the detailed optical observations of both the four dwarfs studied by Davies \& Kinman (1984) and the sample of \citet{tt}. These observations led to the discovery of two other 
very faint UV-emitting objects, possibly members of the group itself (see Fig. \ref{galex_field2} and Table \ref{tab:fos_indb}). All the five dwarfs show NUV-B colors bluer than those of the two brighter members of the group. In particular UGC 2165 is very faint in FUV showing extremely red FUV-NUV colors.

Moreover, with  the aim at matching the specific overall SEDs and the global properties of NGC1023 and NGC1203A in a consistent way with the dynamical properties of the group, we select, from a large set of SPH simulations, one case which gives us insights into the evolution of such a system and  the group itself. 
Such  SPH simulations of galaxy formation and evolution are starting from a collapsing  triaxial systems initially composed of dark matter (DM) and gas in  different proportions and different total masses. These simulations allowed to investigate the DM distribution in the group environment finding that this is not confined to the halo of single galaxies.

The whole SED of NGC~1023 and its global properties are well matched by the  simulation which provides a minor merger, 1:5 $(M)_{sat}/M_{primary}$). The merger phase started 1.8 Gyr before the selected snapshot. 
The whole outcome of the model is in strict agreement with \cite{tt} results: 
their estimate of the the R band absolute magnitude is $M_R=-22.33$ to be compared with our $M_R=-21.95$.
In addition from their study of the NGC 1023 group, the virial mass  inside 315\,Kpc (their virial/harmonic radius), is  6.4$\pm$3$\times$10$^{12}$\, $M_{\odot}$,  and L$_{R}$=2.0$\times$10$^{10}$\, $L_{\odot}$, according to our prediction of a total mass of 5.54$\times$10$^{12}$\, $M_{\odot}$ with a stellar mass of 1.20$\times$10$^{11}$\, $M_{\odot}$,  and L$_{R}$=2.56$\times$10$^{10}$\, $L_{\odot}$ inside the same radius. 

Looking at its evolution, we predict the group more loose and poor, still no-virialized when the final merger phase is not yet begun, 2\, Gyr before our fit, which corresponds to an age of 11.9 Gyr.  B and R luminosities are almost the same as at the age of the fit, but there is 20\% less mass within the same reference radius, i.e. 315\, Kpc \citep{tt}. Furthermore, B and R luminosities reduce by ~1 and  ~0.5 mag respectively 2 Gyr after the age of our fit, the total mass being the same.

\section*{Acknowledgments}

We thank the referee Dr. N. Trentham whose comments greatly improved the paper.
We acknowledge the financial contribution from University of Padova (ex 60\%) and  the agreement ASI-INAF I/009/10/0. GALEX is a NASA Small Explorer, 
launched in April 2003. GALEX is operated for NASA by California Institute of Technology under NASA contract NAS-98034.

\appendix

\end{document}